\documentclass[sigconf]{acmart}

\usepackage{booktabs} 
\usepackage{bbm}
\usepackage{algorithm}
\usepackage{algpseudocode}
\usepackage{graphicx}
\usepackage{subfigure}
\usepackage{amsmath}
\usepackage{booktabs}
\usepackage{amsmath}
\usepackage{tabularx}

\usepackage{array,ragged2e}

\newcolumntype{P}[1]{>{\RaggedRight\hspace{0pt}}p{#1}}

\graphicspath{ {figures/} }

\newsavebox{\tempfig}

\acmDOI{10.475/123_4}

\acmISBN{123-4567-24-567/08/06}

\acmConference[Submisson to KDD'2019]{Submisson to KDD'2019}{August 2019}{Anchorage, Alaska USA}
\acmYear{1997}
\copyrightyear{2016}

\acmArticle{4}
\acmPrice{15.00}

\begin{document}
\title{Behavior Sequence Transformer for E-commerce Recommendation in Alibaba}

\author{Qiwei Chen, Huan Zhao}
\authornote{Qiwei Chen and Huan Zhao contribute equally to this work, and Pipei Huang is the corresponding author.}
\author{Wei Li, Pipei Huang, Wenwu Ou}
\affiliation{%
	\institution{Alibaba Search\&Recommendation Group}
\city{Beijing\&Hangzhou}
\state{China}
}
\email{{chenqiwei.cqw,chuanfei.zh,rob.lw,pipei.hpp,santong.oww}@alibaba-inc.com}

%
%
%
%

\settopmatter{printacmref=false} 
\renewcommand\footnotetextcopyrightpermission[1]{} 
\pagestyle{plain} 

\def\be{\textbf{e}}

\def\bE{\textbf{E}}
\def\bF{\textbf{F}}
\def\bQ{\textbf{Q}}
\def\bK{\textbf{K}}
\def\bV{\textbf{V}}
\def\bS{\textbf{S}}

\def\bW{\textbf{W}}
\def\bv{\textbf{v}}

\def\cV{\mathcal{V}}


\newcommand{\huan}[1]{{\color{blue}{\textbf{Huan: }#1}}}

\begin{abstract}
Deep learning based methods have been widely used in industrial recommendation systems (RSs). Previous works adopt an Embedding\&MLP paradigm: raw features are embedded into low-dimensional vectors, which are then fed on to MLP for final recommendations. However, most of these works just concatenate different features, ignoring the sequential nature of users' behaviors. In this paper, we propose to use the powerful Transformer model to capture the sequential signals underlying users' behavior sequences for recommendation in Alibaba. Experimental results demonstrate the superiority of the proposed model, which is then deployed online at Taobao and obtain significant improvements in online Click-Through-Rate (CTR) comparing to two baselines.

\end{abstract}
\vspace{-0.2in}

\maketitle

\section{Introduction}
\label{sec-intro}
During the last decade, Recommender Systems (RSs) have been the most popular application in industry, and in the past five years, deep learning based methods have been widely used in industrial RSs, e.g., Google~\cite{cheng2016wide,covington2016deep} and Airbnb~\cite{grbovic2018real}. In Alibaba, the largest e-commerce platform in China, RSs have been the key engine for its Gross Merchandise Volume (GMV) and revenues, and various deep learning based recommendation methods have been deployed in rich e-commerce scenarios~\cite{wang2017hybrid,ni2018perceive,zhou2018deep,zhu2018learning,wang2018billion,chen2019pog,li2019multi,pei2019personalized}. As introduced in~\cite{wang2018billion}, the RSs in Alibaba are a two-stage pipeline: match and rank. In match, a set of similar items are selected according to items users interacted with, and then a fine-tuned prediction model is learned to predict the probability of users clicking the given set of candidate items.

In this paper, we focus on the rank stage at Alibaba's Taobao, China's largest Consumer-to-Consumer (C2C) platform owned by Alibaba, where we are have millions of candidate items, and we need to predict the probability of a user clicking the candidate items given his/her historical behaviors. In the era of deep learning, embedding and MLP have been the standard paradigm for industrial RSs: large numbers of raw features are embedded into low-dimensional spaces as vectors, and then fed into fully connected layers, known as multi layer perceptron (MLP), to predict whether a user will click an item or not. The representative works are wide and deep learning (WDL) networks~\cite{cheng2016wide} from Google and Deep Interest networks (DIN) from Alibaba~\cite{zhou2018deep}.

At Taobao, we build the rank model on top of WDL, where various features are used in the embedding\&MLP paradigm , e.g., the category and brand of an item, the statistical numbers of an item, or the user profile features. Despite the success of this framework, it is inherently far from satisfying since it ignores one type of very important signals in practice, i.e., the sequential signal underlying the users' behavior sequences, i.e., users' clicked items in order. In reality, the order matters for predicting the future clicks of users. For example, a user tends to click a case for a cellphone after he or she bought an iphone at Taobao, or tries to find a suitable shoes after buying a pair of trousers. In this sense, it is problematic without considering this factor when deploying a prediction model in the rank stage at Taobao. In WDL~\cite{cheng2016wide}, they simply concatenates all features without capturing the order information among users' behavior sequences. In DIN~\cite{zhou2018deep}, they proposed to use attention mechanism to capture the similarities between the candidate item and the previously clicked items of a user, while it did not consider the sequential nature underlying the user's behavior sequences.

Therefore, in this work, to address the aforementioned problems facing WDL and DIN, we try to incorporate sequential signal of users' behavior sequences into RS at Taobao. Inspired by the great success of the Transformer for machine translation task in natural language processing (NLP)~\cite{vaswani2017attention,devlin2018bert}, we apply the self-attention mechanism to learn a better representation for each item in a user's behavior sequence by considering the sequential information in embedding stage, and then feed them into MLPs to predict users' responses to candidate items. The key advantage of the Transformer is that it can better capture the dependency among words in sentences by the self-attention mechanism, and intuitively speaking, the ``dependency'' among items in users' behavior sequences can also be extracted by the Transformer. Therefore, we propose the user behavior sequence transformer (BST) for e-commerce recommendation at Taobao. Offline experiments and online A/B test show the superiority of BST comparing to existing methods. The BST has been deployed in rank stage for Taobao recommendation, which provides recommending service for hundreds of millions of consumers everyday.

\begin{figure*}
	\includegraphics[width=0.7\textwidth]{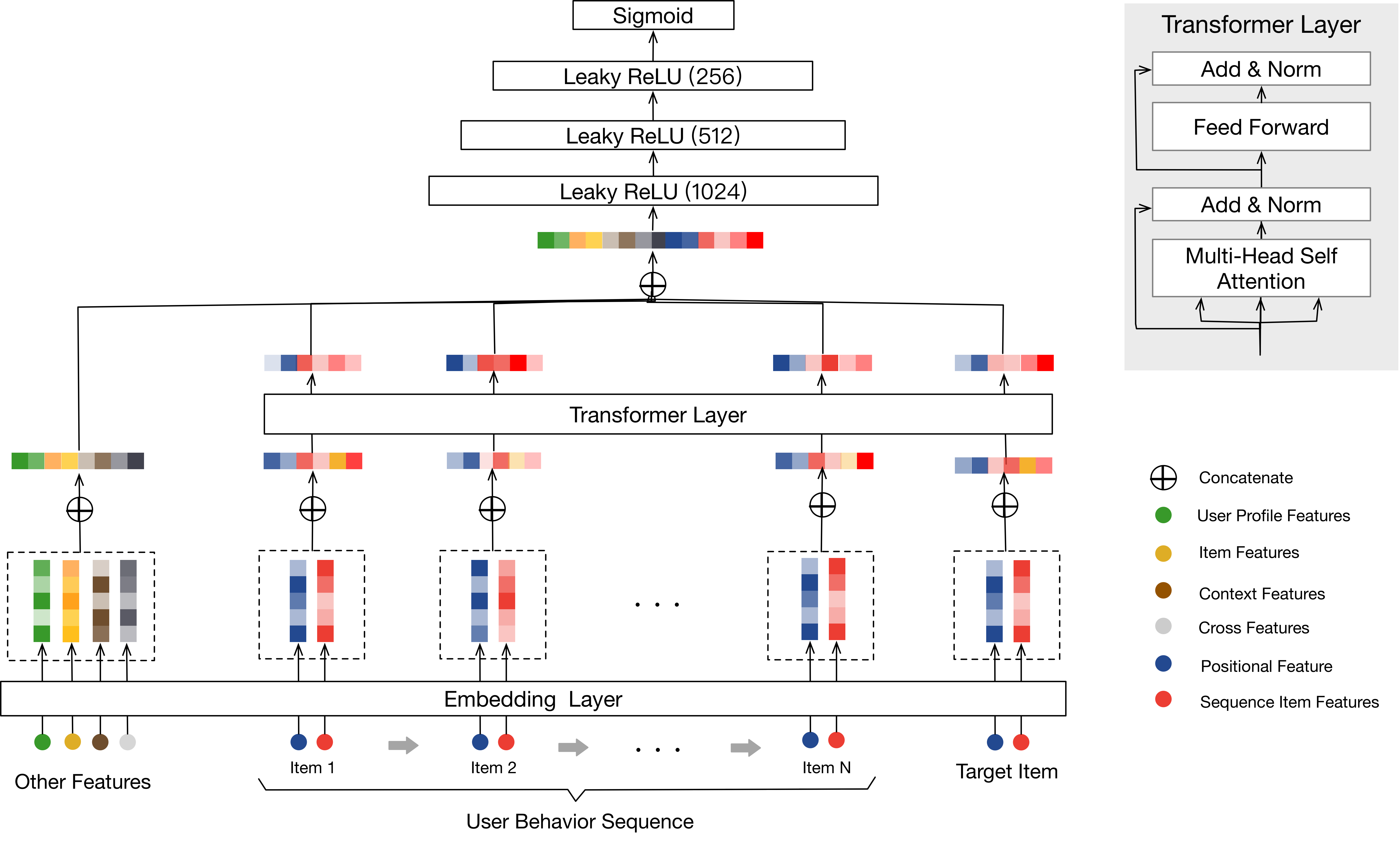}
	\caption{The overview architecture of the proposed BST. BST takes as input the user's behavior sequence, including the target item, and ``Other Features''. It firstly embeds these input features as low-dimensional vectors. To better capture the relations among the items in the behavior sequence, the transformer layer is used to learn deeper representation for each item in the sequence. Then by concatenating the embeddings of Other Features and the output of the transformer layer, the three-layer MLPs are used to learn the interactions of the hidden features, and sigmoid function is used to generate the final output. Note that the ``Positional Features'' are incorporated into ``Sequence Item Features''.}
	\label{fig-architecture}
	\vspace{-0.1in}
\end{figure*}

The remainder of this paper is organized as follows: the architecture is elaborated in Section~\ref{sec-archi}, and then the experimental results including offline and online ones are presented in Section~\ref{sec-exp}. Related work are reviewed in Section~\ref{sec-rel}, and finally we conclude our work in Section~\ref{sec-conclusion}.
\vspace{-0.2in}

\section{Architecture}
\label{sec-archi}
In the rank stage, we model the recommendation task as Click-Through Rate (CTR) prediction problem, which can be defined as follows: given a user's behavior sequence $S(u) = \{v_1, v_2, ..., v_n\}$ clicked by a user $u$, we need to learn a function, $\mathcal{F}$, to predict the probability of $u$ clicking the target item $v_t$, i.e., the candidate one. Other Features include user profile, context, item, and cross features.

We build BST on top of WDL~\cite{cheng2016wide}, and the overview architecture is shown in Figure~\ref{fig-architecture}. From Figure~\ref{fig-architecture}, we can see that it follows the popular embedding\&MLP paradigm, where the previously clicked items and related features are firstly embedded into low-dimensional vectors before fed on to MLP. The key difference between BST and WDL is that we add transformer layer to learn better representations for users' clicked items by capturing the underlying sequential signals. In the following parts, we introduce in a bottom-up manner the key components of BST: embedding layer, transformer layer, and MLP.

\subsection{Embedding Layer}
The first component is the embedding layer, which embeds all input features into a fixed-size low-dimensional vectors. In our scenarios, there are various features, like the user profile features, item features, context features, and the combination of different features, i.e., the cross features\footnote{Though the combination of features can be automatically learned by neural networks, we still incorporate the some hand-crafted cross features, which have been demonstrated useful in our scenarios before the deep learning era.}. Since this work is focused on modeling the behavior sequence with transformer, we denote all these features as ``Other Features'' for simplicity, and give some examples in Table~\ref{tb-features}. As shown in Figure~\ref{fig-architecture}, we concatenate ``Other features'' in left part and embed them into low-dimensional vectors. For these features, we create an embedding matrix $\bW_o \in \mathbb{R}^{|D| \times d_o}$, where $d_o$ is the dimension size.

\begin{table}[]
	\centering
	\caption{The ``Other Features`` shown in left side of Figure~\ref{fig-architecture}. We use much more features in practice, and show a number of effective ones for simplicity.}
	\vspace{-0.1in}
	\label{tb-features}
	\begin{tabular}{cccc}
		\toprule
		User &Item & Context & Cross \\\midrule
		gender &category\_id &match\_type & age * item\_id \\
		age  &shop\_id &display position & os * item\_id \\
		city &tag &page No. & gender * category\_id\\
		...  &... &... & ...\\\bottomrule
	\end{tabular}
\end{table}

Besides, we also obtain the embeddings for each item in the behavior sequence, including the target item. As shown in Figure~\ref{fig-architecture}, we use two types of features to represent an item, ``Sequence Item Features''(in red) and ``Positional Features'' (in dark blue), where ``Sequence Item Features'' include \textit{item\_id} and \textit{category\_id}. Note that an item tends to have hundreds of features, while it is too expensive to choose all to represent the item in a behavior sequence. As introduced in our previous work~\cite{wang2018billion}, the \textit{item\_id} and \textit{category\_id} are good enough for the performance, we choose these two as sparse features to represent each item in embedding the user's behavior sequences. The ``Positional Features'' corresponds the following ``positional embedding''. Then for each item, we concatenate Sequence Item Features and Positional Features, and create an embedding matrix $\bW_V \in \mathbb{R}^{|V| \times d_V}$, where $d_V$ is the dimension size of the embedding, and $|V|$ is the number of items. We use $\be_i \in \mathbb{R}^{d_V}$ to represent the embedding for the $i$-th item in a given behavior sequence.

\textbf{Positional embedding.} In~\cite{vaswani2017attention}, the authors proposed a positional embedding to capture the order information in sentences. Likewise, the order exists in users' behavior sequences. Thus, we add the ``position'' as an input feature of each item in the bottom layer before it is projected as a low-dimensional vector. Note that the position value of item $v_i$ is computed as $pos(v_i) = t(v_t) - t(v_i)$, where $t(v_t)$ represents the recommending time and $t(v_i)$ the timestamp when user click item $v_i$. We adopt this method since in our scenarios it outperforms the $sin$ and $cos$ functions used in~\cite{vaswani2017attention}.
\vspace{-0.1in}

\subsection{Transformer layer}
\label{sec-transformer}
In this part, we introduce the Transformer layer, which learns a deeper representation for each item by capturing the relations with other items in the behavior sequences. 

\textbf{Self-attention layer.} The scaled dot-product attention~\cite{vaswani2017attention} is defined as follows:
\begin{equation}
\text{Attention}(\bQ, \bK, \bV) = \text{softmax}\big(\frac{\bQ\bK^T}{\sqrt{d}}\big)\bV,
\end{equation}
where $\bQ$ represents the queries, $\bK$ the keys and $\bV$ the values. In our scenarios, the self-attention operations takes the embeddings of items as input, and converts them to three matrices through linear projection, and feeds them into an attention layer. Following~\cite{vaswani2017attention}, we use the multi-head attention:
\begin{align}
\bS =& \text{MH}(\bE) = \text{Concat}(head_1, head_2,\cdots,head_h)\bW^H,\\
head_i =&  \text{Attention}(\bE\bW^Q, \bE\bW^K, \bE\bW^V),
\end{align}
where the projection matrices $\bW^Q, \bW^K, \bW^V \in \mathbb{R}^{d \times d}$, and $\bE$ is the embedding matrices of all items, and $h$ is the number of heads.

\textbf{Point-wise Feed-Forward Networks.} Following~\cite{vaswani2017attention}, we add point-wise Feed-Forward Networks (FFN) to further enhance the model with non-linearity, which is defined as follows:
\begin{equation}
\bF = FFN(\bS).
\end{equation}
To avoid overfitting and learn meaningful features hierarchically, we use dropout and LeakyReLU both in self-attention and FFN. Then the overall output of the self-attention and FFN layers are as follows:
\begin{align}
\bS' =& \text{LayerNorm}(\bS + \text{Dropout}(MH(\bS)),\\
\bF  =& \text{LayerNorm}\big(\bS' + \text{Dropout}(\text{LeakyReLU}(\bS'\bW^{(1)} + b^{(1)})\bW^{(2)} + b^{(2)})\big),
\end{align}
where $\bW^{(1)}, b^{(1)}, \bW^{(2)}, b^{(2)}$ are the learnable parameters, and $LayerNorm$ is the standard normalization layer.

\textbf{Stacking the self-attention blocks.} 
After the first self-attention block, it aggregates all the previous items' embeddings, and to further model the complex relations underlying the item sequences, we stack the self-building blocks and the $b$-th block is defined as:
\begin{align}
\bS^b =& SA(F^{(b - 1)}),\\
\bF^b =& FFN(\bS^b), \forall i \in {1,2,\cdots,n}.
\end{align}
In practice, we observe in our experiments $b=1$ obtains better performance comparing to $b=2,3$ (see Table~\ref{tb-res}). For the sake of efficiency, we did not try larger $b$ and leave this for future work.

\subsection{MLP layers and Loss function}
By concatenating the embeddings of Other Features and the output of the Transformer layer applying to the target item, we then use three fully connected layers to further learn the interactions among the dense features, which is standard practice in industrial RSs.

To predict whether a user will click the target item $v_t$, we model it as a binary classification problem, thus we use the sigmoid function as the output unit. To train the model, we use the cross-entropy loss:
\begin{equation}
\mathcal{L} = -\frac{1}{N}\sum_{(x,y) \in \mathcal{D}}\big(y\log p(x) + (1-y)\log(1-p(x))\big),
\end{equation}
where $\mathcal{D}$ represent all the samples, and $y \in \{0, 1\}$ is the label representing whether user have click an item or not, $p(x)$ is the output of the network after the sigmoid unit, representing the predicted probability of sample $x$ being clicked.

\section{Experiments}
\label{sec-exp}
In this section, we present the experimental results.
\vspace{-0.1in}

\begin{table}[]
	\centering
	\caption{Statistics of the constructed Taobao dataset.}
	\vspace{-0.1in}
	\begin{tabular}{c|ccc}
		\toprule
		Dataset & \#Users & \#Items & \#Samples \\\midrule
		Taobao & 298,349,235 & 12,166,060 & 47,556,271,927\\\bottomrule
	\end{tabular}
\label{tb-stat}
\end{table}

\subsection{Settings}
\textbf{Dataset.} The dataset is constructed from the log of Taobao App~\footnote{https://www.taobao.com/}. We construct an offline dataset based on users' behaviors in eight days. We use the first seven days as training data, and the last day as test data. The statistics of the dataset is shown in Table~\ref{tb-stat}. We can see that the dataset is extremely large and sparse.

\noindent\textbf{Baselines.} To show the effectivene of BST, we compare it with two models: WDL~\cite{cheng2016wide} and DIN~\cite{zhou2018deep}. Besides, we create a baseline method by incorporating sequential information into WDL, denoted as WDL(+Seq), which aggregates the embeddings of the previously clicked items in average. Our framework is built on top of WDL by adding sequential modeling with the Transformer, while DIN is proposed to capture the similarities between the target item and the previous clicked items with attention mechanisms.

\noindent\textbf{Evaluation metric.} For the offline results, we use Area Under Curve (AUC) score to evaluate the performance of different models. For online A/B test, we use the CTR and average RT to evaluate all models. RT is short for response time (RT), which is the time cost of generating recommending results for a given query, i.e., one request of a user at Taobao. We use average RT as the metric to evaluate the efficiency of different in online production environment.


\noindent\textbf{Settings.} Our model is implemented with Python 2.7 and Tensorflow 1.4, and the \textit{``Adagrad''} is chosen as the optimizer. Besides, we give the detail of the model parameters in Table~\ref{tb-paras}.
\vspace{-0.1in}

\subsection{Results Analysis}

The results are shown in Table~\ref{tb-res}, from which, we can see the superiority of BST comparing to baselines. In specific, the AUC of offline experiment is improved from $0.7734$ (WDL) and $0.7866$ (DIN) to $0.7894$ (BST). When comparing WDL and WDL(+Seq), we can see that the effectiveness of incorporating sequential information in an simple averaging manner. It means with the help of self-attention, BST provides a powerful capability to capture the sequential signal underlying users' behavior sequences. Note that from our practical experience, even the small gain of offline AUC can lead to huge gain in online CTR. A similar phenomenon is reported by the researchers from Google in WDL~\cite{cheng2016wide}.

Besides, in terms of efficiency, the average RT of BST is close to those of WDL and DIN, which guarantees the feasibility of deploying a complex model like the Transformer in real-world large-scale RSs.

Finally, we also shown the influences of stacking the self-attention layers in Section~\ref{sec-transformer}. From Table~\ref{tb-res}, we can see that $b=1$ obtains the best offline AUC. This may be due to the fact that the sequential dependency in users' behavior sequence is not as complex as that in sentences in machine translation task, thus smaller number of blocks are enough to obtain good performance. Similar observation is reported in~\cite{kang2018self}. Therefore we choose $b=1$ to deploy BST in production environment, and only report the online CTR gain for $b=1$ in Table~\ref{tb-res}.

\begin{table}[]
	\caption{The configuration of BST, and the meaning of the parameters can be inferred from their names.}
	\label{tb-paras}
	\vspace{-0.1in}
	\begin{tabular}{c|c|c|c}
		\toprule
		\multicolumn{4}{c}{Configuration of BST.} \\ \hline
		embedding size & 4 $\sim$64 & batch size & 256 \\ 
		head number & 8 & dropout & 0.2 \\ 
		sequence length & 20 & \#epochs & 1 \\ 
		transformer block & 1 & queue capacity & 1024 \\
		MLP Shape & 1024 * 512 * 256 & learning rate & 0.01 \\ \bottomrule
	\end{tabular}
\end{table}
\vspace{-0.1in}

\section{Related Work}
\label{sec-rel}
In this section, we briefly review the related work on deep learning methods for CTR prediction. Since the proposal of WDL~\cite{cheng2016wide}, a series of works have been proposed to improve the CTR with deep learning based methods, e.g., DeepFM~\cite{guo2017deepfm}, XDeepFM~\cite{lian2018xdeepfm}, Deep and Cross networks~\cite{Wang:2017:DCN:3124749.3124754}, etc. However, all these previous works focus on feature combinations or different architectures of neural networks, ignoring the sequential nature of users' behavior sequence in real-world recommendation scenarios. Recently, DIN~\cite{zhou2018deep} was proposed to deal with users' behavior sequences by an attention mechanism. The key difference between our model and DIN lies in that we propose to use the Transformer~\cite{vaswani2017attention} to learn a deeper representation for each item in users' behavior sequences, while DIN tried to capture different similarities between the previously clicked items and the target item. In other words, our model with transformer are more suitable for capturing the sequential signals. In~\cite{kang2018self,sun2019bert4rec}, the Transformer model is proposed to solve the sequential recommendation problem in sequence-to-sequence manner while the architectures are different from our model in terms of CTR prediction.
\vspace{-0.1in}


\section{Conclusion}
\label{sec-conclusion}
In this paper, we present the technical detail of how we apply the Transformer~\cite{vaswani2017attention} to Taobao recommendation. By using the powerful capability of capturing sequential relations, we show the superiority of the Transformer in modeling user behavior sequences for recommendation by extensive experiments. Besides, we also present the detail of deploying the proposed model in production environment at Taobao, which provides recommendation service for hundreds of millions of users in China.

\begin{table}[]
	\centering
	\caption{Offline AUCs and online CTR gains of different methods. Online CTR gain is relative to the control group.}
	\vspace{-0.1in}
	\begin{tabular}{c|ccc}
		\toprule
		Methods & Offline AUC & Online CTR Gain & Average RT(ms)\\ \midrule
		WDL & 0.7734 & - & 13\\ 
		WDL(+Seq) & 0.7846 & +3.03\% & 14\\ 
		DIN & 0.7866 & +4.55\% & 16\\ \midrule
		BST($b=1$) & \textbf{0.7894} & \textbf{+7.57\%} &20\\ 
		BST($b=2$) & 0.7885 & - &-\\
		BST($b=3$) & 0.7823 & - &-\\
		\bottomrule
	\end{tabular}
	\label{tb-res}
\end{table}

\section{Acknowledgments}
We would like to thank colleagues of our team - Jizhe Wang, Chao Li, Zhiyuan Liu, Yuchi Xu and Mengmeng Wu for useful discussions and supports on this work. We are grateful to Jinbin Liu, Shanshan Hao and Yanchun Yang from Alibaba Distributed Computing Team, and Kan Liu, Tao Lan from Alibaba Online Inference Team, who help deploy the model in production. We also thank the anonymous reviewers for their valuable comments and suggestions that help improve the quality of this manuscript.

\bibliographystyle{ACM-Reference-Format}
\bibliography{ref}
\end{document}